\documentclass[%
 aip,
 jmp,%
 amsmath,amssymb,
preprint,
]{revtex4-1}
\usepackage{amsfonts}
\usepackage{url}
\usepackage{graphicx}
\usepackage[all]{xy} 

\newtheorem{thm}{Theorem}[subsection]

\newtheorem{lem}[thm]{Lemma}

\def\t{\mathfrak{t}}
\def\C{\mathbb{C}}
\def\R {\mathbb{R}}
\def\M {\mathcal{M}}
\def\B {\mathcal{B}}
\def\H {\mathcal{H}}
\def\N {\mathcal{N}}

\begin{document}


\title{Dimension of Conformal Blocks in Five Dimensional K\"{a}hler-Chern-Simons Theory} 



\author{Haitao Liu}
  \email{haitao.liu@unb.ca}
\affiliation{ 
Department of Mathematics and Statistics,\\
  University of New Brunswick, Fredericton, Canada, E3B 5A3
}%



\date{\today}

\begin{abstract}
We briefly review the  K\"{a}hler-Chern-Simon theory on 5-manifolds which are trivial circle bundles over 4-dimensional K\"{a}hler manifolds and present a detailed calculation of the path integral, using the method of Blau and Thompson.
\end{abstract}


\maketitle 

\section{Introduction}

Conformal Field Theories (CFTs) are powerful tools for investigating string theory and
for
addressing certain mathematical questions, for example, mirror pairs.  
However the technology of CFTs derives from 2D quantum field theory, and
this limits applicability to situations that are, at some level, two
dimensional.

There is evidence \cite{Polchinski1995, Polchinski1996} that theories involving objects  with two or more
degrees of freedom might be important for a full understanding of gravity and
elementary particles.  Thus it would seem reasonable to examine theories which are
higher dimensional versions of CFTs.  However, in general such theories do not have
nearly as rich of a structure as 2D CFTs, and this is because the conformal group in
three or more dimensions is finite dimensional, in contrast to the infinite dimensional
Kac-Moody group which arises in 2D CFTs.  

However there are higher dimensional theories that possess many of the important
characteristics of 2D CFTs.  These theories are not conformally invariant in the usual
sense, but they are associated with infinite dimensional algebras that strongly
resemble, indeed, generalize the Kac-Moody groups \cite{Billig1998, Losev1995, Kac1994}.  The most
directly relevant of these theories are the WZW$_4$ models \cite{Losev1996}.  However, these
are in turn associated (in the same manner that WZW$_2$ models are associated with
Chern-Simons gauge theories in 3D) with higher dimensional gauge theories, namely 5D
Chern -Simons and K\"{a}hler-Chern-Simons theories \cite{Zanelli:1999fs, Gegenberg2000a, Nair1990}.  What is especially intriguing
here is that two dimensional integrable models can be described by these theories.

In the beautiful paper by Losev et. al. \cite{Losev1996}, WZW$_4$ models were fairly exhaustively
studied. However, the closely related KCS models have not been as thoroughly studied. 
In this note, and in further work in progress, I intend to explore more deeply the
quantum aspects of KCS theories.

Suppose that $P$ is a trivial principle $SU(r+1)$ bundle over a four dimensional K\"{a}hler manifold $(M,\omega)$, where $\omega$ is the K\"{a}hler form and $A$ is an arbitrary connection on $P$. In \cite{Nair1990, Nair1992} Nair and Schiff introduced the so-called K\"{a}hler-Chern-Simons action:
\begin{equation}
S=\frac{1}{4\pi}\int_{M\times \R} \omega\wedge Tr[CS(A)]+dt\wedge Tr[(\Phi^{2,0}\wedge F+\Phi^{0,2}\wedge F)], 
\label{1}
\end{equation}
where $CS(A)=A\wedge dA+\frac{2}{3}A\wedge A\wedge A$ is the Chern-Simons 3-form, $t$ is the "time" coordinate on $\R$, $\Phi^{2,0}$ and $\Phi^{0,2}$ are two Lagrange multipliers that are Lie algebra valued $(2,0)$ and $(0,2)$ forms on $M$ respectively and $F$ is the curvature 2-form corresponding to a connection $A$. According to geometric quantization the physical Hilbert space, which is called the space of \emph{conformal blocks}, should be $\mathcal{H}=H^0(\M, \mathcal{L})$ where $\M$ is the phase space and $\mathcal{L}$ is the prequantum line bundle \cite{Woodhouse1997}.  In \cite{Losev1996} the author presented a formula for the dimension of the conformal blocks of the K\"{a}hler-Chern-Simons theory. The aim of this paper is to derive this formula by using Blau and Thompson's method \cite{Blau1993b}. 

In section 2 we will analyze this system at the classical level. We show that the phase space is just the moduli space of Anti-Self-Dual(ASD) instantons. In section 3 and the appendix we will calculate the partition function of the K\"{a}hler-Chern-Simons theory on $M\times S^1$ based on the diagonalization assumption.

\section{Classical K\"{a}hler-Chern-Simons Theory}

First let us split the connection $A$ into spatial and "time" parts $A=A_0+B$ where $A_0$ is a Lie-algebra valued 1-form on $\R$ and $B$ is a Lie-algebra valued 1-form on $M$. Substituting $A=A_0+B$ into the K\"{a}hler-Chern-Simons action (\ref{1}) we get 
\begin{equation}
S=\frac{1}{4\pi}\int_{M\times \R} \omega\wedge Tr[2A_0\wedge (dB+B\wedge B)+B\wedge d_0 B]+ dt\wedge Tr(\Phi^{2,0}\wedge F_B^{0,2}+\Phi^{0,2}\wedge F_B^{2,0}),
\label{7}
\end{equation}
where $d$ and $d_0$ are the differential operators on $M$ and $\R$, respectively, and $F^{0,2}(F^{2,0})$ is the $(0,2)((2,0))$ part of the curvature 2-form of connection $B$ on $M$. Varying the multipliers $\Phi^{2,0}$ and $\Phi^{0,2}$ in the K\"{a}hler-Chern-Simons action (\ref{1}) gives equations
\begin{eqnarray}
F_B^{0,2}=F_B^{2,0}=0.
\label{3}
\end{eqnarray}
Further, varying $A_0$ in the first term in (\ref{7}) gives 
\begin{equation}
\omega\wedge F_B=0.
\label{4}
\end{equation}
These are the equations of motion.
Then the phase space is the moduli space $\mathcal{M}$ of the solutions of (\ref{3},\ref{4}), which is the moduli space of ASD instantons. On $\mathcal{M}$ there is a natural symplectic form 
\begin{equation}
\Omega(a,b):=\frac{1}{2\pi}\int_M \omega\wedge a\wedge b.
\label{6}
\end{equation}
According to geometric quantization, the physical Hilbert space, which is called the space of \emph{conformal blocks}, should be $\mathcal{H}=H^0(\M, \mathcal{L})$ where $\M$ is the phase space and $\mathcal{L}$ is the prequantum line bundle. Further, the Donaldson-Yau-Uhlenbeck theorem tells us that $\M$ is equivalent to the moduli space of semi-stable bundles \cite{Donaldson1997}. As with three dimensional Chern-Simons theory \cite{Blau1993b}, we can borrow the statistical mechanics formula 
\begin{equation}
Z_{M\times S^1}=\textrm{Tr}e^{-\beta H},
\end{equation}
for a circle of radius $\beta$ to calculate the dimension of the physical Hilbert space. From now on we only consider the K\"{a}hler-Chern-Simons theory on $M\times S^1$.

\section{Quantization}
According to the geometric quantization program\cite{Losev1996, Woodhouse1997, Losev1997} the physical Hilbert space $\H$ should be the space of sections of the Quillen determinant bundle \cite{Bismut1988a, Bismut1988b, Bismut1988c, Losev1997}. Our purpose is to try to calculate $\textrm{dim}\H$. 
\subsection{Gauge Fixing}
From now on we will focus on the following action and add the constraints (\ref{3}) on the $B$ fields. In next subsection, I will give the definition of the path integral over the constraint surface by using Henneaux and Teitelboim's formula \cite{Henneaux1994}. Hence we write
\begin{eqnarray}
S&=&\frac{1}{4\pi}\int_{M\times S^1} \omega\wedge Tr[2A_0\wedge (dB+B\wedge B)+B\wedge d_0 B],\nonumber \\
 &=& \frac{1}{4\pi}\int_{M\times S^1} \omega\wedge Tr[2A_0\wedge dB+B\wedge D_0 B],
 \label{8}
\end{eqnarray}
where $D_0:=d_0+[A_0, \cdot ]$ is the covariant derivative operator on $S^1$. Since the Lie algebra $\mathfrak{su}(r+1)$ has the orthogonal decomposition $\mathfrak{g}=\t\oplus k$, the gauge fields $A_0$ and $B$ have the following decomposition: 
\begin{eqnarray}
  A_0&=&A_0^{\t}+A_0^{k}, \label{9}\\
      B&=& B^{\t}+B^{k}.  \label{10}
\end{eqnarray}
Subtituting eq. (\ref{9}, \ref{10}) into the action (\ref{8}) we get 
\begin{equation}
S=\frac{1}{4\pi}\int_{M\times S^1} \omega\wedge Tr[2A_0^{\t}\wedge dB^{\t}+2A_0^k\wedge dB^k+B^{\t}\wedge d_0 B^{\t}+B^k\wedge D_0 B^k ].
\label{11}
\end{equation}

 According to \cite{Blau1993b} we can choose the following gauge fixing 
 \begin{eqnarray}
\partial_t A_{0t}&=&0, \label{12}\\
A_0^k&=&0. \label{13}
\end{eqnarray}
Here $A_{0t}$ are the components of $A_0$, i.e. $A_{0}=A_{0t}dt$. This was suggested by the corresponding insertion for the 3-dimemsional Chern-Simons theory \cite{Blau1993b}. Thus we insert the following term into the action:
\begin{equation}
\frac{1}{4\pi}\int_{M\times S^1} \omega \wedge Tr(\Psi \wedge A_0+2i\omega\wedge \bar{c}D_0c) \label{14}
\end{equation}
where $c,\bar{c}$ are Grassmanian-valued functions and $\Psi$ is a Lie algebra valued 2-form on $M$. The constraints on $c, \bar{c}$ and $\Psi$ are 
\begin{equation}
\oint_{S^1}c^{\t}=\oint_{S^1}\bar{c}^{\t}=\oint_{S^1} \Psi^{\t}\wedge dt=0.   \label{15}
\end{equation}
The first term of eq. (\ref{14}) is equivalent to adding the constraint (\ref{13}). 
 Then after choosing the gauge fixing the action becomes
 \begin{eqnarray}
 S=\frac{1}{4\pi}\int_{M\times S^1} \omega\wedge Tr(2A_0^{\t}\wedge dB^{\t}&+&B^{\t}\wedge d_0 B^{\t}+B^k\wedge D_0 B^k \nonumber\\
     &&+2i\omega\wedge \bar{c}^{\t}D_0c^{\t}+2i\omega\wedge \bar{c}^{k}D_0c^{k}),
 \label{16}
 \end{eqnarray}
 where the $B$ fields satisfy the constraints (\ref{3}) and the $A_0$ fields satisfy the constraint (\ref{13}). Before entering the next subsection, let us analyze the gauge fixing carefully. We notice that in the action (\ref{16}) there are still residual gauge symmetries 
 \begin{eqnarray}
 A_0^{\t}&\rightarrow& A_0^{\t}, \nonumber \\
 B&\rightarrow& g^{-1}Bg+g^{-1}dg, 
 \end{eqnarray}
 where $g$ is a map from $M\times S^1$ to the maximum torus T in $SU(r+1)$ satisfying $\partial_t g=0$. Later we will use this residual gauge symmetry to regularize the functional determinant. 
 
Since $M$ is a K\"{a}hler manifold, the 1-forms $B^{\t}$ and $B^k$ have the following decomposition:
\begin{eqnarray}
B^{\t}&=&\sum_{i=1}^2(B^{\t}_i\phi^i+B^{\t}_{\bar{i}}\phi^{\bar{i}}), \label{18}\\
B^k&=&\sum_{i=1}^2(B^k_i\phi^i+B^k_{\bar{i}}\phi^{\bar{i}}), \label{19}
\end{eqnarray}
where $\phi^i, \phi^{\bar{i}}$ are the holomorphic and antiholomorphic 1-form fields respectively, and satisfy $\omega=\frac{i}{2}\sum_{i=1}^2\phi^{i}\wedge \phi^{\bar{i}}$ \cite{P.Griffiths1994}. Plugging (\ref{18}) and (\ref{19}) into the action S (\ref{16}), we get 
\begin{eqnarray}
S=&&\frac{1}{2\pi}\int_{M\times S^1} \omega\wedge dt \wedge Tr(A_{0t}^{\t}\wedge dB^{\t})+idV\wedge dt Tr(\sum_iB^{\t}_i \partial_t B^{\t}_{\bar{i}}\nonumber\\
    && +\sum_iB^k_i \nabla_t B^k_{\bar{i}}+\bar{c}^{\t}\nabla_tc^{\t}+\bar{c}^k\nabla_tc^k), 
	\label{20} 
\end{eqnarray}
where $dV=\omega\wedge \omega$ is the volume form of $M$ and $\nabla_t=\frac{d}{dt}+[A^{\t}_{0t},.]$. $B^{\t}$ is independent of time. This can be shown by using the Fourier expansion with respect to the time circle $S^1$: 
\begin{equation}
B^{\t}=B_0^{\t}e^{2n\pi it}.
\end{equation}
Thus 
\begin{equation}
\int_{M\times S^1} \omega\wedge dt \wedge Tr(A_{0t}^{\t}\wedge dB^{\t})=\int_{M\times S^1} \omega\wedge dt \wedge Tr(A_{0t}^{\t}\wedge dB_0^{\t}).
\end{equation}
Here we already use the gauge fixing condition eq. (\ref{12}). In other words, the $B^{\t}$ in term $\int_{M\times S^1} \omega\wedge dt \wedge Tr(A_{0t}^{\t}\wedge dB^{\t})$ only contains the time independent part. 

\subsection{Path Integral}

We know that Blau and Thompson's method has not been established in higher dimensions \cite{Blau1995}. So here for simplicity we assume the diagnalization is applicable and assume that the complete set of obstructions to diagonalizing are all the $T-$bundle restrictions of the trivial $SU(r+1)$ bundle \cite{Blau1995}, where $T$ is the maximum torus of $SU(r+1)$. So after gauge fixing, the partition function $Z$ is defined as
\begin{equation}
 Z:=\int DA_{0t}^{\t}D\B^{\t}Dc^{\t}D\bar{c}^{\t}Dc^kD\bar{c}^k\prod_{i=1}^2DB^{\t}_i D\bar{B}^{\t}_i DB^k_i D\bar{B}^k_i \sqrt{Det\Omega} e^{iS},
\label{17} 
 \end{equation}
where the $B, \bar{B}$ fields are on the constraint surface $F^{(0,2)}=F^{(2,0)}=0$. 
In order to avoid symbol confusion we use $\B^{\t}$ to replace $B^{\t}$. According to the analysis of last section, $\B^{\t}$ is the time independent part of the full $B^{\t}$. Here we borrow Henneaux and Teitelboim's path integral definition on the constraint surface \cite{Henneaux1994}. We know in eq. (\ref{1}) the equations $F^{(0,2)}=F^{(2,0)}=0$ are the second class constraints \cite{Nair1990}. So according to Henneaux and Teitelboim's definition we need to insert $\sqrt{Det\Omega}$ as the determinant of the symplectic form on the constraint surface.
Following Henneaux and Teitelboim's book \cite{Henneaux1994}, it is easy to show that this definition (\ref{17}) is equivalent to 
\begin{eqnarray}
 Z:=&&\int DA_{0t}^{\t}D\B^{\t}Dc^{\t}D\bar{c}^{\t}Dc^kD\bar{c}^k\prod_{i=1}^2DB^{\t}_i D\bar{B}^{\t}_i DB^k_i D\bar{B}^k_i \cdot \nonumber\\ 
&&\cdot \sqrt{Det\{F^{(2,0)}, F^{(0,2)}\}_{P.B.}} \delta(F^{(2,0)})\delta(F^{(0,2)})e^{iS},  \label{70} 
 \end{eqnarray}
where $\{F^{(2,0)}, F^{(0,2)}\}_{P.B.}$ is the Poinsson bracket between $F^{(2,0)}$ and $F^{(0,2)}$, $\delta(F^{(2,0)}),\delta(F^{(0,2)})$ are the Dirac delta functions coming from the path integral over the Lagrangian multipliers $\Phi, \bar{\Phi}$, and $B, \bar{B}$ fields are unconstraint.

Now we do the "background expansion". First choose a background $B_c^{\t}, \bar{B}_c^{\t}$ which is time independent and satisfy $F_{\bar{B}_c^{\t}}^{(0,2)}=F_{B_c^{\t}}^{(2,0)}=0$. Then expand a connection in the small neighborhood of $B_c^{\t}, \bar{B}_c^{\t}$ on the constraint surface as 
\begin{eqnarray}
B=B_{c}^{\t}+B_{q}^{\t}+B_{q}^{k}, \nonumber\\
\bar{B}=\bar{B}_{c}^{\t}+\bar{B}_{q}^{\t}+\bar{B}_{q}^{k} \label{65}, 
\label{eq:61}
\end{eqnarray}
where $B_q^{\t},\bar{B}_q^{\t}$ is time dependent. In another word, for  an arbitrary connection $B$ using the Fourier expansion with respect to the time circle $S^1$, we have 
\begin{equation}
B=B^{\t}_{n}e^{2n\pi it}+B_{n}^ke^{2n\pi it}.
\label{eq:}
\end{equation}
The background expansion means that $B_c^{\t}=B_0^{\t}\ll B-B_0^{\t}$ and $B_c^{\t}=B_0^{\t}$ is on the constraint surface. Hence, we have the following equations for $B_c^{\t}, \bar{B}_c^{\t}, B_q^{\t},\bar{B}_q^{\t}, B_{q}^k, \bar{B}_q^k$:
$$
\left\{ \begin{array}{cccccc}
          \partial B_{c}^{\t}=\bar{\partial}\bar{B}_{c}^{\t}=\partial B_{q}^{\t}=\bar{\partial}\bar{B}_{q}^{\t}=0 \\
          \partial_{B_{c}^{\t}}B_{q}^k=\bar{\partial}_{\bar{B}_c^{\t}}\bar{B}_{q}^k=0 .
          \end{array} \right .
$$
Thus subtituting the background expansion (\ref{eq:61}) into the action eq. (\ref{20}), we get 
\begin{eqnarray}
S=&&\frac{1}{2\pi}\int_{M\times S^1} \omega\wedge dt \wedge Tr(A_{0t}^{\t}\wedge d\B_c^{\t})+idV\wedge dt Tr(\sum_i B^{\t}_{qi} \partial_t \bar{B}^{\t}_{qi}\nonumber\\
    && +\sum_iB_{qi}^k \nabla_t \bar{B}^k_{qi}+\bar{c}^{\t}\nabla_tc^{\t}+\bar{c}^k\nabla_tc^k), 
	\label{64} 
\end{eqnarray}
We define $Z_{\B_c}$ as
\begin{equation}
Z_{\B_c^{\t}}:=\int Dc^{\t}D\bar{c}^{\t}Dc^kD\bar{c}^k\prod_{i=1}^2DB^{\t}_{qi} D\bar{B}^{\t}_{qi} DB^k_{qi} D\bar{B}^k_{qi} \sqrt{Det\Omega_{\B_c^{\t}}} e^{iS}.
\end{equation}
Here we expand $\sqrt{Det\Omega}$ around $\B_c^{\t}=B_c^{\t}+\bar{B}_c^{\t}$ and denote the leading term as $\sqrt{Det\Omega_{\B_c^{\t}}}$.
Hence our partition function eq. (\ref{17}) is 
\begin{equation}
Z=\int DA_{0t}^{\t}D\B_c^{\t} Z_{\B_c^{\t}}
\end{equation}

 So after integrating out all the modes of $B^{\t}_{qi}, B^{\t}_{q\bar{i}}, B^k_{qi},B^k_{q\bar{i}}, c^{\t},\bar{c}^{\t}, c^k, \bar{c}^k$, the partition function becomes 
\begin{eqnarray}
Z=\int DA_{0t}^{\t}D\B_c^{\t} \sqrt{Det\Omega_{\B_c^{\t}}} &&\frac{Det^{'}_{\t}(\partial_{t})_{\Omega^{0,0}(M)\otimes \Omega^0(S^1)}}{Det^{'}_{\t}(\partial_{t})_{\Omega^{0,1*}(M)\otimes \Omega^0(S^1)}}\cdot\frac{Det_{k}(\nabla_{t})_{\Omega^{0,0}(M)\otimes \Omega^0(S^1)}}{Det_{k}(\nabla_{t})_{\Omega^{0,1*}(M)\otimes \Omega^0(S^1)}}\cdot \nonumber \\
 && \cdot\exp[\frac{i}{2\pi}\int_M \omega\wedge Tr(A^{\t}_{0t}d\B_c^{\t})],
\label{21}
\end{eqnarray}
where $Det^{'}$ has no $S^1$ zero modes and $\Omega^{0,1*}_{\t}(M), \Omega^{0,1*}_{k}(M)$ are defined as follows:
\begin{eqnarray}
\Omega^{0,1*}_{\t}(M)&=&\{ \bar{B}_q\in \Omega^{0,1}(M;\t)|\bar{\partial}\bar{B}_q^{\t}=0\}, \label{22}\\
\Omega^{0,1*}_{k}(M)&=&\{\bar{B}_q\in \Omega^{0,1}(M;k)|\bar{\partial}\bar{B}_q^k+\bar{B}_c^{\t}\wedge \bar{B}_q^k=0\}. \label{23}
\end{eqnarray}

\subsection{Evaluation Of The Abelianized Partition Function}

After substituting the values of the above functional determinants into the path integral (In the Appendix we show how to evaluate these functional determinants), we get 
\begin{eqnarray}
Z&=& \int DA_0D\B_c \sqrt{Det\Omega_{\B_c}} e^{\frac{ih}{2\pi}\int_M Tr(A_0F_{\B_c})c_1(M) }\times (-1)^{\frac{1}{\pi}\int_M\rho(F_{\B_c})c_1(M) } \cdot\nonumber\\
  && \ \ \ \cdot\prod_{\alpha>0}\Big(2\sin \frac{\alpha(A_0)}{2}\Big)^{\frac{1}{6}(c_1^2(M)+c_2(M))+c_1^2(k_{\alpha})} \cdot\exp \Big\{\frac{i}{2\pi}\int_M \omega\wedge Tr(A_0F_{\B_c})\} \nonumber\\
  &=&\int DA_0D\B \sqrt{Det\Omega_{\B_c}} e^{\frac{i}{2\pi}\int_M Tr(A_0F_{\B_c})\wedge (\omega+hc_1(M)) }\times (-1)^{\frac{1}{\pi}\int_M \rho (F_{\B_c})c_1(M) } \cdot\nonumber\\
  && \ \ \ \cdot\prod_{\alpha>0}\Big(2\sin \frac{\alpha(A_0)}{2}\Big)^{\frac{1}{6}(c_1^2(M)+c_2(M))+c_1^2(k_{\alpha})}.
 \label{60}
\end{eqnarray}

According to \cite{Blau1993b}, we know after fixing the gauge, $D\B_c=DF_{\B_c}$. So eq. (\ref{60}) becomes
\begin{eqnarray}
Z&=&\int DA_0DF_{\B_c}\sqrt{Det\Omega_{\B_c}} e^{\frac{i}{2\pi}\int_M Tr(A_0F_{\B_c})\wedge (\omega+hc_1(M)) }\times (-1)^{\frac{1}{\pi}\int_M \rho (F_{\B_c})c_1(M) } \cdot\nonumber\\
  && \ \ \ \cdot\prod_{\alpha>0}\Big(2\sin \frac{\alpha(A_0)}{2}\Big)^{\frac{1}{6}(c_1^2(M)+c_2(M))+c_1^2(k_{\alpha})}
 \label{61}
\end{eqnarray}

Denote $\phi=A_0$, $[F_{\B_c}]=\eta\in H^2(M, \mathbb{Z}^r)$ where $\eta=(\eta_{\alpha_1}, \cdots, \eta_{\alpha_{r}})\in \textrm{Pic}(M)^r$. Then the above equation can be rewritten as 

\begin{eqnarray}
Z&=&\sum_{\substack{
       \eta\in \textrm{Pic}(M)^r \\
     ch_2(\oplus \eta_{\alpha})=0
	 }}
	\sqrt{Det\Omega_{\eta}} \int_{\t\cap \triangle_+} D\phi e^{\frac{i}{2\pi}\int_M Tr(\phi[\eta])\wedge (\omega+hc_1(M)) }\times (-1)^{\frac{1}{\pi}\int_M \rho ([\eta])c_1(M) } \cdot\nonumber\\
  && \ \ \ \cdot\prod_{\alpha>0}\Big(2\sin \frac{\alpha(A_0)}{2}\Big)^{\frac{1}{6}(c_1^2(M)+c_2(M))+[\eta]_{\alpha}\cdot[\eta]_{\alpha}}
 \label{62}
\end{eqnarray}
where $\t\cap \triangle_+$ is the Weyl alcove \cite{Blau1993b, Losev1996} and $ch_2(\oplus \eta_{\alpha})$ \cite{Blau1995} is the 2nd Chern character of bundle $\oplus \eta_{\alpha}$. The condition $ch_2(\oplus \eta_{\alpha})=0$ comes from the assumption
	that the complete set of obstructions to diagonalizing are all the
	$T-$bundle restrictions of the trivial $SU(r+1)$ bundle, where $T$
	is the maximum torus of $SU(r+1)$. In \cite{Blau1995} Blau and Thompson
	discussed the condition for the existence of this restriction.

\section{Conclusion And Further Study}
As in the three dimensional SU(N) Chern-Simons theory, the partition function $Z$ is equal to the dimension of the physical Hilbert space, up to a renormalization \cite{Blau1993b}, i.e.
\begin{eqnarray}
\textrm{dim}\mathcal{H} &=& N\cdot Z=N \sum_{\substack{
       \eta\in \textrm{Pic}(M)^r \\
     ch_2(\oplus \eta_{\alpha})=0
	 }}
	\sqrt{Det\Omega_{\eta}} \int_{\t\cap \triangle_+} D\phi e^{\frac{i}{2\pi}\int_M Tr(\phi[\eta])\wedge (\omega+hc_1(M)) }\times (-1)^{\frac{1}{\pi}\int_M \rho ([\eta])c_1(M) } \cdot\nonumber\\
  && \ \ \ \cdot\prod_{\alpha>0}\Big(2\sin \frac{\alpha(A_0)}{2}\Big)^{\frac{1}{6}(c_1^2(M)+c_2(M))+[\eta]_{\alpha}\cdot[\eta]_{\alpha}}
\label{63}
\end{eqnarray}
The factor $\sqrt{Det\Omega_{\eta}}$ is compatible with \cite{Hyun:1995is,Losev1996, Baulieu:1997nj} in the following sense: in \cite{Hyun:1995is,Losev1996,Baulieu:1997nj} the authors used the BRST (or equivariant cohomology) technique. In their results the path integral 
\begin{equation}
\int D\Psi D\bar{\Psi} \exp\{i\int_{X^4\times S^1} Tr(\omega \wedge dt\wedge\Psi\wedge\bar{\Psi})\}
\end{equation}
corresponds to our factor $\sqrt{Det\Omega_{\eta}}$. The difference is that our factor $\sqrt{Det\Omega_{\eta}}$ is evaluated at certain "points" on the reduced surface. Further, it remains to determine the normalization $N$. For three dimensional Chern-Simons theory we can borrow the formula for the volume of the flat connection moduli space \cite{Blau1993b,Witten1991a}. This suggests an investigation of the K\"{a}hler-Chern-Simons theory to see whether we can use the volume of the ASD connection moduli space \cite{Moore2000,Witten1994} to calculate the normalization $N$. Work on this is in progress.

\begin{acknowledgments}
I would like to thank J. Gegenberg and C. Ingalls for valuable discussion on this project. 
\end{acknowledgments}

\appendix
\section{APPENDIX: Evluating The Functional Determinants}
In this subsection we will focus on the calculation of the determinants 
\begin{equation}
\frac{Det^{'}_{\t}(\partial_{t})_{\Omega^{0,0}(M)\otimes \Omega^0(S^1)}}{Det^{'}_{\t}(\partial_{t})_{\Omega^{0,1*}(M)\otimes \Omega^0(S^1)}},
\label{24}
\end{equation}
and 
\begin{equation}
\frac{Det_{k}(\nabla_{t})_{\Omega^{0,0}(M)\otimes \Omega^0(S^1)}}{Det_{k}(\nabla_{t})_{\Omega^{0,1*}(M)\otimes \Omega^0(S^1)}}. \label{25}
\end{equation}
In order to make sense of these determinants, we need to regularize them. In Section 2 we already discussed the residual Abelian gauge symmetry. That implies that the regularization should not break the residual gauge symmetry. We will use the heat kernel regularization based on the $\t$ covariant Laplacian $\triangle_{\B_c}=-(\partial_{\B_c}^*\partial_{\B_c}+\partial_{\B_c}\partial_{\B_c}^*)$  \cite{Blau1993b}. Here we will use the same definition of the determinant as \cite{Blau1993b,Blau2006}:

For an operator $\mathcal{O}$ we define
\begin{equation}
\log Det \mathcal{O}:= Tr (e^{-\epsilon\triangle_{\B_c}}\log \mathcal{O})
\label{26}
\end{equation}
Before calculating the determinants (\ref{24}) and (\ref{25}), let us analyze some exact sequences. 

$P$ is a trivial $SU(r+1)$ bundle on $M\times S^1$ and $adP$ denotes its adjoint bundle which is still trivial. It is easy to find that the pullback bundles on $M$ should also be trivial. From now on we denote $P$ and $adP$ as pullback bundles on $M$. 
\begin{lem}\label{lem1}
If $H^{0,1}(M, \mathbb{C})=H^{0,2}(M, \mathbb{C})=0$, then $H^{0,1}(M, adP\otimes \C)=H^{0,2}(M, adP\otimes \C)=0$.
\end{lem}
If $\t$ is a subbundle of $adP\otimes \C$, then $adP\otimes \C$ decomposes as $adP\otimes \C=\t\oplus k$ under the natural metric "Tr" of $adP\otimes \C$. In other words, we have the following exact sequence:
\begin{equation}
\xymatrix{
0 \ar[r] &\t \ar[r] & adP\otimes \C \ar[r] & k \ar[r] & 0 }. 
\end{equation}
Then we get the following long exact sequence:
\begin{equation}
\xymatrix{
\ar[r] & H^{0,1}(M, \t) \ar[r] &H^{0,1}(M, adP\otimes \C)\ar[r] & H^{0,1}(M,k) \ar[r] &\\
\ar[r] & H^{0,2}(M, \t) \ar[r] &H^{0,2}(M, adP\otimes \C)\ar[r] & H^{0,2}(M,k) \ar[r] &0}
\end{equation}
From the Lemma\ref{lem1} we get 
\begin{eqnarray}
H^{0,2}(M, k)&=&0,  \\
H^{0,1}(M, k) &\cong& H^{0,2}(M, \t).
\end{eqnarray}
Similarly, using the exact sequence
\begin{equation}
\xymatrix{
0 \ar[r] &k \ar[r] & adP\otimes \C \ar[r] & \t \ar[r] & 0 }, 
\end{equation}
we can show that 
\begin{eqnarray}
H^{0,2}(M, \t)&=&0,  \\
H^{0,1}(M, \t) &\cong& H^{0,2}(M, k).
\end{eqnarray}
So we have 
\begin{equation}
H^{0,2}(M, k)=H^{0,1}(M, k)=H^{0,1}(M, \t)=H^{0,2}(M, \t)=0. \label{27}
\end{equation}
Now for the $\bar{B}_c^{\t}$ satisfying $\bar{\partial}\bar{B}_c^{\t}=0$, we have the following complex
\begin{equation}
\xymatrix{
0 \ar[r] &\Omega^{0,0}(M)\otimes\t  \ar[r]^{\bar{\partial}_{\bar{B}_c^{\t}}}& \Omega^{0,1}(M)\otimes \t \ar[r]^{\bar{\partial}_{\bar{B}_c^{\t}}} & \Omega^{0,2}(M)\otimes\t \ar[r] & 0 }. \label{28}
\end{equation}
For $\forall f\in \Omega^{0,0}(M)\otimes\t$ and $\forall \alpha\in\Omega^{0,1}(M)\otimes \t$, 
\begin{eqnarray}
\bar{\partial}_{\bar{B}_c^{\t}}f&=&\bar{\partial}f+\bar{B}_c^{\t}f,   \label{29}\\
\bar{\partial}_{\bar{B}_c^{\t}}\alpha &=&\bar{\partial}\alpha+\bar{B}_c^{\t}\alpha.
\label{30}
\end{eqnarray}
Similarly for $B_c^{t}$, which is the complex conjugate of $\bar{B}_c^{\t}$, we have 
\begin{equation}
\xymatrix{
0 \ar[r] &\Omega^{0,0}(M)\otimes\t  \ar[r]^{\partial_{B_c^{\t}}}& \Omega^{1,0}(M)\otimes \t \ar[r]^{\partial_{B_c^{\t}}} & \Omega^{2,0}(M)\otimes\t \ar[r] & 0 }. \label{31}
\end{equation}
Thus $\forall h\in \Omega^{0,0}(M)\otimes$ and $\forall \beta\in\Omega^{1,0}(M)\otimes \t$, 
\begin{eqnarray}
\partial_{B_c^{\t}}h&=&\partial h+B_c^{\t}h,   \label{32} \\
\partial_{B_c^{\t}}\beta &=&\partial\beta+B_c^{\t}\beta.   
\label{33}
\end{eqnarray}
Denote 
\begin{eqnarray}
\B_c^t&:=&B_c^{\t}+\bar{B}_c^{\t}, \\
d_{\B_c^{\t}}&:=&\partial_{B_c^{\t}}+\bar{\partial}_{\bar{B}_c^{\t}}.
\end{eqnarray}
Then for $\forall f\in \Omega^0(M)\otimes \t$, we have 
Next we will investigate the determinant (\ref{23}).
According to the regularization (\ref{26}) we have
\begin{eqnarray}
\log \frac{Det^{'}_{\t}(\partial_{t})_{\Omega^{0,0}(M)\otimes \Omega^0(S^1)}}{Det^{'}_{\t}(\partial_{t})_{\Omega^{0,1*}(M)\otimes \Omega^0(S^1)}}&=& (Tr_{\Omega^{0,0}_{\t}(M)}e^{-\epsilon\triangle_{\bar{B}_c^{\t}}}-Tr_{\Omega^{0,1*}_{\t}(M)}e^{-\epsilon\triangle_{\bar{B}_c^{\t}}})\log Det^{'}_{\t}\partial_t|_{\Omega^0(S^1)} \nonumber\\
&=& (\textrm{dim} H^{0,0}(M, \t)-\textrm{dim} H^{0,1}(M, \t))\log Det^{'}_{\t}\partial_t|_{\Omega^0(S^1)} \nonumber \\
&=& \textrm{Ind} \bar{\partial}_{\bar{B}_c^{\t}} \log Det^{'}_{\t}\partial_t|_{\Omega^0(S^1)} \nonumber\\
&=& \int_M ch(\t)Td(M) \log Det^{'}_{\t}\partial_t|_{\Omega^0(S^1)} \nonumber \\
&=& \int_M [r+c_1(\t)+\frac{1}{2}c_1^2(\t)-c_2(\t)][1+c_1(M)+\frac{1}{12}(c_1^2(M)+c_2(M))]\cdot \nonumber \\
  &&\ \ \ \ \ \cdot \log Det^{'}_{\t}\partial_t|_{\Omega^0(S^1)} \nonumber\\
&=& \int_M [r\frac{1}{12}(c_1^2(M)+c_2(M))]\cdot \log Det^{'}_{\t}\partial_t|_{\Omega^0(S^1)},
\label{35}
\end{eqnarray}
where $\textrm{Ind} \bar{\partial}_{\bar{B}_c^{\t}}$ denotes the index of operator $\bar{\partial}_{\bar{B}_c^{\t}}$ for the complex (\ref{28}), $ch(\t)$ denotes the Chern character of bundle $\t$, $Td(M)$ denotes the Todd genus of M and $r$ denote the rank of bundle $\t$. During the calculation we have used the index theorem. Furthermore, according to \cite{Blau1993b, Witten1991a}, we know, up to a normalization,  
\begin{equation}
Det^{'}_{\t}\partial_t|_{\Omega^0(S^1)}\sim 1. \label{36}
\end{equation}

Then we have 
\begin{equation}
\frac{Det^{'}_{\t}(\partial_{t})_{\Omega^{0,0}(M)\otimes \Omega^0(S^1)}}{Det^{'}_{\t}(\partial_{t})_{\Omega^{0,1*}(M)\otimes \Omega^0(S^1)}}\sim 1. \label{37}
\end{equation}
Next we will calculate the determinant (\ref{25}). We know the bundle $k$ has decomposition $k=\oplus_{\alpha>0}(k_{\alpha}\oplus k_{-\alpha})$ where $\alpha>0$ are the positive roots of the Lie algebra $\mathfrak{su}(r+1)$. Then we have the exact sequence 
\begin{equation}
\xymatrix{
0 \ar[r] & k_{\alpha} \ar[r] & k \ar[r] &\N \ar[r] & 0}, \label{47}
\end{equation}
where $\N$ is the normal bundle of $k_{\alpha}$. 
Then we get the long exact sequence 
\begin{equation}
\xymatrix{
\ar[r] & H^{0,1}(M, k_{\alpha}) \ar[r] &H^{0,1}(M, k)\ar[r] & H^{0,1}(M,\N) \ar[r] & \\
\ar[r] & H^{0,2}(M, k_{\alpha}) \ar[r] &H^{0,2}(M, k)\ar[r] & H^{0,2}(M,\N) \ar[r] &0.} 
\end{equation}
According to eq. (\ref{27}) we have 
\begin{eqnarray}
H^{0,2}(M,\N)&=&0, \\
H^{0,1}(M,\N) &\cong&H^{0,2}(M, k_{\alpha}). 
\end{eqnarray}
Similarly using the exact sequence 
\begin{equation}
\xymatrix{
0 \ar[r] &\N \ar[r] & k \ar[r] & k_{\alpha} \ar[r] & 0}, \label{38}
\end{equation}
we can get 
\begin{eqnarray}
H^{0,2}(M,k_{\alpha})&=&0, \\
H^{0,1}(M,k_{\alpha}) &\cong&H^{0,2}(M, \N). 
\end{eqnarray}
So 
\begin{equation}
H^{0,1}(M,k_{\alpha}) =H^{0,2}(M,k_{\alpha})=H^{0,1}(M,\N) =H^{0,2}(M, \N) =0 \label{39}
\end{equation}
Now we have the complex
\begin{equation}
\xymatrix{
0 \ar[r] & \Omega^{0,0}(M)\otimes k_{\alpha} \ar[r]^{\bar{\partial}_{\bar{B}_c^{\t}}} & \Omega^{0,1}(M)\otimes k_{\alpha} \ar[r]^{\bar{\partial}_{\bar{B}_c^{\t}}} & \Omega^{0,2}(M)\otimes k_{\alpha} \ar[r] &0}, \label{40}
\end{equation}
where $\bar{\partial}_{\bar{B}_c^{\t}}$ is same as the previous one in (\ref{28}).
For this complex, according to the index theorem, we have 
\begin{eqnarray}
\textrm{Ind} \bar{\partial}_{\bar{B}_c^{\t}}|_{k_{\alpha}}&=& \textrm{dim} H^{0,0}(M, k_{\alpha})- \textrm{dim} H^{0,1}(M,k_{\alpha})+\textrm{dim} H^{0,2}(M, k_{\alpha})\nonumber\\
&=& \int_M ch(k_{\alpha})Td(M) \nonumber\\
&=& \int_M [1+c_1(k_{\alpha})+\frac{1}{2}c_1^2(k_{\alpha})][1+c_1(M)+\frac{1}{12}(c_1^2(M)+c_2(M))] \nonumber\\
&=& c_1(k_{\alpha})c_1(M)+\frac{1}{12}(c_1^2(M)+c_2(M))+\frac{1}{2}c_1^2(k_{\alpha}). \label{41}
\end{eqnarray}
Now let us derive the $c_1(k_{\alpha})$. We have the following sequence
\begin{equation}
\xymatrix{
0 \ar[r] &\Omega^0(M)\otimes k_{\alpha} \ar[r]^{d_{\B_c^{\t}}} &\Omega^{1}(M)\otimes k_{\alpha} \ar[r]^{d_{\B_c^{\t}}} &\Omega^2(M)\otimes k_{\alpha} \ar[r] & \cdots },\label{42}
\end{equation}
where $d_{\mathcal{B}_c^{\t}}:=\partial_{B_c^{\t}}+\bar{\partial}_{\bar{B}_c^{\t}}$ and $B_c^{\t}$ is the complex conjugate of $\bar{B}_c^{\t}$ and for $\forall f\otimes k_{\alpha}\in \Omega^0(M)\otimes k_{\alpha}$,
\begin{eqnarray}
d_{\B_c^{\t}}(f\otimes k_{\alpha})&=&df\otimes k_{\alpha}+f\otimes [\B_c^{\t}, k_{\alpha}]\nonumber\\
                                  &=& df\otimes k_{\alpha}+f\otimes [\B_{ic}^{\t}h^{i}, k_{\alpha}]\nonumber\\
								  &=&[df+i\B_{ic}^{\t}\alpha^if]\otimes k_{\alpha}, 
\end{eqnarray}
where $\B_{ic}^{\t}=B_{ic}^{\t}+\bar{B}_{ic}^{\t}$ and $h^i, i=1,\cdots, r$ are the basis of $\t$ satisfying $[h^{i}, k_{\alpha}]=i\alpha^ik_{\alpha}$.
So we have 
\begin{equation}
c_1(k_{\alpha})=-\frac{1}{2\pi}\alpha(F_{\B_c^{\t}})=-\frac{1}{2\pi}F_{\B_c^{\t} i}\alpha(h^i)=-\frac{1}{2\pi}F_{\B_c^{\t} i}\alpha^i,      \label{54}
\end{equation}

As in \cite{Blau1993b}, we have
\begin{eqnarray}
\log \frac{Det_{k_{\alpha}}(\nabla_{t})_{\Omega^{0,0}(M)\otimes \Omega^0(S^1)}}{Det_{k_{\alpha}}(\nabla_{t})_{\Omega^{0,1*}(M)\otimes \Omega^0(S^1)}} &=& \Big(Tr_{\Omega^{0,0}_{k_{-\alpha}(M)}}e^{-\epsilon\triangle_{\bar{B}_c^{\t}}}-Tr_{\Omega^{0,1*}_{k_{-\alpha}(M)}}e^{-\epsilon\triangle_{\bar{B}_c^{\t}}}\Big)\cdot \log Det_{k_{\alpha}}\nabla_t|_{\Omega^0(S^1)} \nonumber \\
&=& \Big(\textrm{dim}H^{0,0}(M, k_{-\alpha})-\textrm{dim}H^{0,1}(M,k_{-\alpha})\Big)\cdot \log Det_{k_{\alpha}}\nabla_t|_{\Omega^0(S^1)} \nonumber \\
&=& \textrm{Ind} \bar{\partial}_{\bar{B}_c^{\t}}|_{k_{-\alpha}}\cdot \log Det_{k_{\alpha}}\nabla_t|_{\Omega^0(S^1)} \nonumber \\
&=& \Big[c_1(k_{-\alpha})c_1(M)+\frac{1}{12}(c_1^2(M)+c_2(M))+\frac{1}{2}c_1^2(k_{-\alpha})\Big]\cdot \nonumber \\
 && \ \ \ \ \ \ \ \ \ \ \cdot\log Det_{k_{\alpha}}\nabla_t|_{\Omega^0(S^1)}\nonumber \\
 &=& \Big[-c_1(k_{\alpha})c_1(M)+\frac{1}{12}(c_1^2(M)+c_2(M))+\frac{1}{2}c_1^2(k_{\alpha})\Big]\cdot \nonumber \\
 && \ \ \ \ \ \ \ \ \ \ \cdot\log Det_{k_{\alpha}}\nabla_t|_{\Omega^0(S^1)} . 
 \label{43}
\end{eqnarray}
And similarly we  have 
\begin{eqnarray}
\log \frac{Det_{k_{-\alpha}}(\nabla_{t})_{\Omega^{0,0}(M)\otimes \Omega^0(S^1)}}{Det_{k_{-\alpha}}(\nabla_{t})_{\Omega^{0,1*}(M)\otimes \Omega^0(S^1)}} &=& \Big(Tr_{\Omega^{0,0}_{k_{\alpha}(M)}}e^{-\epsilon\triangle_{\bar{B}_c^{\t}}}-Tr_{\Omega^{0,1*}_{k_{\alpha}(M)}}e^{-\epsilon\triangle_{\bar{B}_c^{\t}}}\Big)\cdot \log Det_{k_{-\alpha}}\nabla_t|_{\Omega^0(S^1)} \nonumber \\
&=&\Big(\textrm{dim}H^{0,0}(M, k_{\alpha})-\textrm{dim}H^{0,1}(M,k_{\alpha})\Big)\cdot \log Det_{k_{-\alpha}}\nabla_t|_{\Omega^0(S^1)} \nonumber \\
&=& \textrm{Ind} \bar{\partial}_{\bar{B}_c^{\t}}|_{k_{\alpha}}\cdot \log Det_{k_{-\alpha}}\nabla_t|_{\Omega^0(S^1)} \nonumber \\
&=& [c_1(k_{\alpha})c_1(M)+\frac{1}{12}(c_1^2(M)+c_2(M))+\frac{1}{2}c_1^2(k_{\alpha})]\cdot \nonumber \\
 && \ \ \ \ \ \ \ \ \ \ \cdot\log Det_{k_{-\alpha}}\nabla_t|_{\Omega^0(S^1)}. 
 \label{44}
\end{eqnarray}
So
\begin{eqnarray}
\log \frac{Det_{k}(\nabla_{t})_{\Omega^{0,0}(M)\otimes \Omega^0(S^1)}}{Det_{k}(\nabla_{t})_{\Omega^{0,1*}(M)\otimes \Omega^0(S^1)}} &=& \sum_{\alpha>0}\Big\{[\frac{1}{12}(c_1^2(M)+c_2(M))+\frac{1}{2}c_1^2(k_{\alpha})]\log Det_{k_{\alpha}}\nabla_t|_{\Omega^0(S^1)} \cdot\nonumber \\
&&\ \ \ \ \ \cdot\log Det_{k_{-\alpha}}\nabla_t|_{\Omega^0(S^1)} - c_1(k_{\alpha})c_1(M) \log \frac{Det_{k_{\alpha}}\nabla_t|_{\Omega^0(S^1)}}{Det_{k_{-\alpha}}\nabla_t|_{\Omega^0(S^1)}} \Big\}. \nonumber\\
&& 
\label{45}
\end{eqnarray}
Finally 
\begin{eqnarray}
\frac{Det_{k}(\nabla_{t})_{\Omega^{0,0}(M)\otimes \Omega^0(S^1)}}{Det_{k}(\nabla_{t})_{\Omega^{0,1*}(M)\otimes \Omega^0(S^1)}} =
&& \prod_{\alpha>0}\exp\Big\{-c_1(k_{\alpha})c_1(M) \log \frac{Det_{k_{\alpha}}\nabla_t|_{\Omega^0(S^1)}}{Det_{k_{-\alpha}}\nabla_t|_{\Omega^0(S^1)}}\Big\} \cdot \nonumber\\
&& \cdot \prod_{\alpha>0} \Big(Det_{k_{\alpha}}\nabla_t|_{\Omega^0(S^1)}Det_{k_{-\alpha}}\nabla_t|_{\Omega^0(S^1)}\Big)^{\frac{1}{12}(c_1^2(M)+c_2(M))+\frac{1}{2}c_1^2(k_{\alpha})}.  \nonumber\\
&&    
\label{46}
\end{eqnarray}

From \cite{Blau1993b, Witten1991a} we know, up to a normalization, 
\begin{eqnarray}
\textrm{Det}_{k_{\alpha}}\nabla_t|_{\Omega^0(S^1)}&\sim& \textrm{Det}[\textrm{Id}-\textrm{Ad}_{k_{\alpha}}(\exp iA_0)]=\textrm{Det} [\textrm{Id}-\textrm{Ad}_{k_{\alpha}}(e^{ i\alpha(A_0)})], \label{48} \\
\textrm{Det}_{k_{-\alpha}}\nabla_t|_{\Omega^0(S^1)}&\sim& \textrm{Det}[\textrm{Id}-\textrm{Ad}_{k_{-\alpha}}(\exp A_0)]=\textrm{Det} [\textrm{Id}-\textrm{Ad}_{k_{-\alpha}}(e^{-i\alpha(A_0)})].  \label{49} 
\end{eqnarray}
We denote 
\begin{eqnarray}
M_{\alpha}&:=&\textrm{Id}-\textrm{Ad}_{k_{\alpha}}(e^{i\alpha(A_0)}), \label{50}\\
M_{-\alpha}&:=&\textrm{Id}-\textrm{Ad}_{k_{-\alpha}}(e^{-i\alpha(A_0)}).  \label{51}
\end{eqnarray}
According to \cite{Blau1993b},
\begin{equation}
\frac{M_{\alpha}}{M_{-\alpha}}=\frac{1-e^{i\alpha(A_0)}}{1-e^{-i\alpha(A_0)}} =-e^{i\alpha(A_0)}.
\label{52}
\end{equation}
So 
\begin{eqnarray}
\prod_{\alpha>0}\exp\Big\{-c_1(k_{\alpha})c_1(M) \log \frac{Det_{k_{\alpha}}\nabla_t|_{\Omega^0(S^1)}}{Det_{k_{-\alpha}}\nabla_t|_{\Omega^0(S^1)}}\Big\}&=& \prod_{\alpha>0}\exp\Big\{-c_1(k_{\alpha})c_1(M) \log \frac{M_{\alpha}}{M_{-\alpha}}\Big\} \nonumber\\
&=& \prod_{\alpha>0}\exp\Big\{-c_1(k_{\alpha})c_1(M) \log (-e^{i\alpha(A_0)})\Big\} \nonumber\\
&=& \prod_{\alpha>0}\Big\{e^{-i\alpha(A_0)c_1(k_{\alpha})c_1(M) } \times (-1)^ {-c_1(k_{\alpha})c_1(M) } \Big\}  \nonumber\\
&=& e^{\sum_{\alpha>0}\Big\{-i\alpha(A_0)c_1(k_{\alpha})c_1(M) \Big\}} \times (-1)^ {\sum_{\alpha>0}\Big\{-c_1(k_{\alpha})c_1(M)  \Big\} }. \nonumber\\
&& 
\label{53}
\end{eqnarray}
According to (\ref{54}) we know
\begin{equation}
-\sum_{\alpha>0}\alpha(A_0)c_1(k_{\alpha})=\sum_{\alpha>0}\frac{1}{2\pi}\int_M \alpha(A_0)\alpha(F_{\B_c})=\frac{h}{2\pi}\int_MTr(A_0F_{\B_c}), \label{55}
\end{equation}
where $h$ is the Coxeter number of $SU(r+1)$ \cite{Blau1993b}.
So 
\begin{equation}
\prod_{\alpha>0}\exp\Big\{-c_1(k_{\alpha})c_1(M) \log \frac{Det_{k_{\alpha}}\nabla_t|_{\Omega^0(S^1)}}{Det_{k_{-\alpha}}\nabla_t|_{\Omega^0(S^1)}}\Big\}=e^{\frac{ih}{2\pi}\int_M Tr(A_0F_{\B_c})c_1(M) }\times (-1)^{\frac{1}{\pi}\int_M \rho (F_{\B_c})c_1(M) }, \label{56}
\end{equation}
where $\rho=\frac{1}{2}\sum_{\alpha>0}\alpha$. 
Finally from \cite{Blau1993b} we know that 
\begin{equation}
Det_{k_{\alpha}}\nabla_t|_{\Omega^0(S^1)}Det_{k_{-\alpha}}\nabla_t|_{\Omega^0(S^1)}=4\sin^2\frac{\alpha(A_0)}{2}. \label{57}
\end{equation}
So we have 
\begin{eqnarray}
&&\prod_{\alpha>0} \Big(Det_{k_{\alpha}}\nabla_t|_{\Omega^0(S^1)}Det_{k_{-\alpha}}\nabla_t|_{\Omega^0(S^1)}\Big)^{\frac{1}{12}(c_1^2(M)+c_2(M))+\frac{1}{2}c_1^2(k_{\alpha})} \nonumber\\
&&=\prod_{\alpha>0}\Big(2\sin \frac{\alpha(A_0)}{2}\Big)^{\frac{1}{6}(c_1^2(M)+c_2(M))+c_1^2(k_{\alpha})}.    
\label{58}
\end{eqnarray}
At last, we have
\begin{eqnarray}
\frac{Det_{k}(\nabla_{t})_{\Omega^{0,0}(M)\otimes \Omega^0(S^1)}}{Det_{k}(\nabla_{t})_{\Omega^{0,1*}(M)\otimes \Omega^0(S^1)}}=&& e^{\frac{ih}{2\pi}Tr(A_0F_{\B_c})c_1(M) }\times (-1)^{\frac{1}{\pi}\int_M\rho(F_{\B_c})c_1(M) } \cdot\nonumber\\
&&\prod_{\alpha>0}\Big(2\sin \frac{\alpha(A_0)}{2}\Big)^{\frac{1}{6}(c_1^2(M)+c_2(M))+c_1^2(k_{\alpha})}. 
\label{59}
\end{eqnarray}


%
%

%



\begin{thebibliography}{10}%
\makeatletter
\providecommand \@ifxundefined [1]{%
 \ifx #1\undefined \expandafter \@firstoftwo
 \else \expandafter \@secondoftwo
\fi
}%
\providecommand \@ifnum [1]{%
 \ifnum #1\expandafter \@firstoftwo
 \else \expandafter \@secondoftwo
\fi
}%
\providecommand \enquote [1]{``#1''}%
\providecommand \bibnamefont  [1]{#1}%
\providecommand \bibfnamefont [1]{#1}%
\providecommand \citenamefont [1]{#1}%
\providecommand\href[0]{\@sanitize\@href}%
\providecommand\@href[1]{\endgroup\@@startlink{#1}\endgroup\@@href}%
\providecommand\@@href[1]{#1\@@endlink}%
\providecommand \@sanitize [0]{\begingroup\catcode`\&12\catcode`\#12\relax}%
\@ifxundefined \pdfoutput {\@firstoftwo}{%
 \@ifnum{\z@=\pdfoutput}{\@firstoftwo}{\@secondoftwo}%
}{%
 \providecommand\@@startlink[1]{\leavevmode}%
 \providecommand\@@endlink[0]{}%
}{%
 \providecommand\@@startlink[1]{%
  \leavevmode
  \pdfstartlink
   attr{/Border[0 0 1 ]/H/I/C[0 1 1]}%
   user{/Subtype/Link/A<</Type/Action/S/URI/URI(#1)>>}%
  \relax
 }%
 \providecommand\@@endlink[0]{\pdfendlink}%
}%
\providecommand \url  [0]{\begingroup\@sanitize \@url }%
\providecommand \@url [1]{\endgroup\@href {#1}{\urlprefix}}%
\providecommand \urlprefix [0]{URL }%
\providecommand \Eprint[0]{\href }%
\@ifxundefined \urlstyle {%
  \providecommand \doi [1]{doi:\discretionary{}{}{}#1}%
}{%
  \providecommand \doi [0]{doi:\discretionary{}{}{}\begingroup
  \urlstyle{rm}\Url }%
}%
\providecommand \doibase [0]{http://dx.doi.org/}%
\providecommand \Doi[1]{\href{\doibase#1}}%
\providecommand \selectlanguage [0]{\@gobble}%
\providecommand \bibinfo [0]{\@secondoftwo}%
\providecommand \bibfield [0]{\@secondoftwo}%
\providecommand \translation [1]{[#1]}%
\providecommand \BibitemOpen[0]{}%
\providecommand \bibitemStop [0]{}%
\providecommand \bibitemNoStop [0]{.\EOS\space}%
\providecommand \EOS [0]{\spacefactor3000\relax}%
\providecommand \BibitemShut [1]{\csname bibitem#1\endcsname}%
\bibitem{Polchinski1995}%
  \BibitemOpen
  \bibfield{author}{%
  \bibinfo {author} {\bibfnamefont{Joseph}\ \bibnamefont{Polchinski}},\ }%
  \bibfield{title}{%
  \enquote{\bibinfo {title} {Dirichlet-branes and ramond-ramond charges},}\ }%
  \bibfield{journal}{%
  \Doi{10.1103/PhysRevLett.75.4724}{\bibinfo {journal} {Phys. Rev. Lett.}}\ }%
  \textbf{\bibinfo {volume} {75}},\ \bibinfo {pages} {4724--4727} (\bibinfo
  {year} {1995})\BibitemShut{NoStop}%
\bibitem{Polchinski1996}%
  \BibitemOpen
  \bibfield{author}{%
  \bibinfo {author} {\bibfnamefont{Joseph}\ \bibnamefont{Polchinski}}, \bibinfo
  {author} {\bibfnamefont{Shyamoli}\ \bibnamefont{Chaudhuri}},\ and\ \bibinfo
  {author} {\bibfnamefont{Clifford~V.}\ \bibnamefont{Johnson}},\ }%
  \bibfield{title}{%
  \enquote{\bibinfo {title} {Notes on d-branes},}\ }%
  \bibfield{journal}{%
  \bibinfo {journal} {hep-th/9602052}}%
   (\bibinfo {year} {1996})\BibitemShut{NoStop}%
\bibitem{Billig1998}%
  \BibitemOpen
  \bibfield{author}{%
  \bibinfo {author} {\bibfnamefont{Yuly}\ \bibnamefont{Billig}},\ }%
  \bibfield{title}{%
  \enquote{\bibinfo {title} {Principal vertex operator representations for
  toroidal lie algebras},}\ }%
  \bibfield{journal}{%
  \Doi{10.1063/1.532472}{\bibinfo {journal} {J. Math. Phys.}}\ }%
  \textbf{\bibinfo {volume} {39}},\ \bibinfo {pages} {3844--3864} (\bibinfo
  {year} {1998})\BibitemShut{NoStop}%
\bibitem{Losev1995}%
  \BibitemOpen
  \bibfield{author}{%
  \bibinfo {author} {\bibfnamefont{Andrei}\ \bibnamefont{Losev}}, \bibinfo
  {author} {\bibfnamefont{Gregory~W.}\ \bibnamefont{Moore}}, \bibinfo {author}
  {\bibfnamefont{Nikita}\ \bibnamefont{Nekrasov}},\ and\ \bibinfo {author}
  {\bibfnamefont{Samson~L.}\ \bibnamefont{Shatashvili}},\ }%
  \bibfield{title}{%
  \enquote{\bibinfo {title} {Central extensions of gauge groups revisited},}\
  }%
  \bibfield{journal}{%
  \bibinfo {journal} {hep-th/9511185}}%
   (\bibinfo {year} {1995})\BibitemShut{NoStop}%
\bibitem{Kac1994}%
  \BibitemOpen
  \bibfield{author}{%
  \bibinfo {author} {\bibfnamefont{Victor~G.}\ \bibnamefont{Kac}},\ }%
  \emph{\bibinfo {title} {Infinite-Dimensional Lie Algebras}},\ \bibinfo
  {edition} {3rd}\ ed.\ (\bibinfo {publisher} {Cambridge University Press},\
  \bibinfo {year} {1994})\BibitemShut{NoStop}%
\bibitem{Losev1996}%
  \BibitemOpen
  \bibfield{author}{%
  \bibinfo {author} {\bibfnamefont{Andrei}\ \bibnamefont{Losev}}, \bibinfo
  {author} {\bibfnamefont{Gregory~W.}\ \bibnamefont{Moore}}, \bibinfo {author}
  {\bibfnamefont{Nikita}\ \bibnamefont{Nekrasov}},\ and\ \bibinfo {author}
  {\bibfnamefont{Samson}\ \bibnamefont{Shatashvili}},\ }%
  \bibfield{title}{%
  \enquote{\bibinfo {title} {Four-dimensional avatars of two-dimensional
  rcft},}\ }%
  \bibfield{journal}{%
  \Doi{10.1016/0920-5632(96)00015-1}{\bibinfo {journal} {Nucl. Phys. Proc.
  Suppl.}}\ }%
  \textbf{\bibinfo {volume} {46}},\ \bibinfo {pages} {130--145} (\bibinfo
  {year} {1996})\BibitemShut{NoStop}%
\bibitem{Zanelli:1999fs}%
  \BibitemOpen
  \bibfield{author}{%
  \bibinfo {author} {\bibfnamefont{Jorge}\ \bibnamefont{Zanelli}},\ }%
  \bibfield{title}{%
  \enquote{\bibinfo {title} {Chern-simons gravity: From 2+1 to 2n+1
  dimensions},}\ }%
  \bibfield{journal}{%
  \bibinfo {journal} {Braz. J. Phys.}\ }%
  \textbf{\bibinfo {volume} {30}},\ \bibinfo {pages} {251--267} (\bibinfo
  {year} {2000})\BibitemShut{NoStop}%
\bibitem{Gegenberg2000a}%
  \BibitemOpen
  \bibfield{author}{%
  \bibinfo {author} {\bibfnamefont{J.}~\bibnamefont{Gegenberg}}\ and\ \bibinfo
  {author} {\bibfnamefont{G.}~\bibnamefont{Kunstatter}},\ }%
  \bibfield{title}{%
  \enquote{\bibinfo {title} {Boundary dynamics of higher dimensional
  chern-simons gravity},}\ }%
  \bibfield{journal}{%
  \bibinfo {journal} {hep-th/0010020}}%
   (\bibinfo {year} {2000})\BibitemShut{NoStop}%
\bibitem{Nair1990}%
  \BibitemOpen
  \bibfield{author}{%
  \bibinfo {author} {\bibfnamefont{V.~P.}\ \bibnamefont{Nair}}\ and\ \bibinfo
  {author} {\bibfnamefont{Jeremy}\ \bibnamefont{Schiff}},\ }%
  \bibfield{title}{%
  \enquote{\bibinfo {title} {A kahler-chern-simons theory and quantization of
  instanton moduli spaces},}\ }%
  \bibfield{journal}{%
  \Doi{10.1016/0370-2693(90)90624-F}{\bibinfo {journal} {Phys. Lett.}}\ }%
  \textbf{\bibinfo {volume} {B246}},\ \bibinfo {pages} {423--429} (\bibinfo
  {year} {1990})\BibitemShut{NoStop}%
\bibitem{Nair1992}%
  \BibitemOpen
  \bibfield{author}{%
  \bibinfo {author} {\bibfnamefont{V.~P.}\ \bibnamefont{Nair}}\ and\ \bibinfo
  {author} {\bibfnamefont{Jeremy}\ \bibnamefont{Schiff}},\ }%
  \bibfield{title}{%
  \enquote{\bibinfo {title} {Kahler chern-simons theory and symmetries of
  antiselfdual gauge fields},}\ }%
  \bibfield{journal}{%
  \Doi{10.1016/0550-3213(92)90239-8}{\bibinfo {journal} {Nucl. Phys.}}\ }%
  \textbf{\bibinfo {volume} {B371}},\ \bibinfo {pages} {329--352} (\bibinfo
  {year} {1992})\BibitemShut{NoStop}%
\bibitem{Woodhouse1997}%
  \BibitemOpen
  \bibfield{author}{%
  \bibinfo {author} {\bibfnamefont{N.~M.~J.}\ \bibnamefont{Woodhouse}},\ }%
  \emph{\bibinfo {title} {Geometric Quantization}},\ Oxford Mathematical
  Monographs\ (\bibinfo {publisher} {Oxford University Press},\ \bibinfo {year}
  {1997})\BibitemShut{NoStop}%
\bibitem{Blau1993b}%
  \BibitemOpen
  \bibfield{author}{%
  \bibinfo {author} {\bibfnamefont{Matthias}\ \bibnamefont{Blau}}\ and\
  \bibinfo {author} {\bibfnamefont{George}\ \bibnamefont{Thompson}},\ }%
  \bibfield{title}{%
  \enquote{\bibinfo {title} {Derivation of the verlinde formula from
  chern-simons theory and the g/g model},}\ }%
  \bibfield{journal}{%
  \Doi{10.1016/0550-3213(93)90538-Z}{\bibinfo {journal} {Nucl. Phys.}}\ }%
  \textbf{\bibinfo {volume} {B408}},\ \bibinfo {pages} {345--390} (\bibinfo
  {year} {1993})\BibitemShut{NoStop}%
\bibitem{Donaldson1997}%
  \BibitemOpen
  \bibfield{author}{%
  \bibinfo {author} {\bibfnamefont{S.~K.}\ \bibnamefont{Donaldson}}\ and\
  \bibinfo {author} {\bibfnamefont{P.~B.}\ \bibnamefont{Kronheimer}},\ }%
  \emph{\bibinfo {title} {The Geometry of Four-Manifolds}}\ (\bibinfo
  {publisher} {Oxford University Press},\ \bibinfo {year}
  {1997})\BibitemShut{NoStop}%
\bibitem{Losev1997}%
  \BibitemOpen
  \bibfield{author}{%
  \bibinfo {author} {\bibfnamefont{Andrei}\ \bibnamefont{Losev}}, \bibinfo
  {author} {\bibfnamefont{Gregory~W.}\ \bibnamefont{Moore}}, \bibinfo {author}
  {\bibfnamefont{Nikita}\ \bibnamefont{Nekrasov}},\ and\ \bibinfo {author}
  {\bibfnamefont{Samson}\ \bibnamefont{Shatashvili}},\ }%
  \bibfield{title}{%
  \enquote{\bibinfo {title} {Chiral lagrangians, anomalies, supersymmetry, and
  holomorphy},}\ }%
  \bibfield{journal}{%
  \Doi{10.1016/S0550-3213(96)00612-8}{\bibinfo {journal} {Nucl. Phys.}}\ }%
  \textbf{\bibinfo {volume} {B484}},\ \bibinfo {pages} {196--222} (\bibinfo
  {year} {1997})\BibitemShut{NoStop}%
\bibitem{Bismut1988a}%
  \BibitemOpen
  \bibfield{author}{%
  \bibinfo {author} {\bibfnamefont{J.~M.}\ \bibnamefont{Bismut}}, \bibinfo
  {author} {\bibfnamefont{H.}~\bibnamefont{Gillet}},\ and\ \bibinfo {author}
  {\bibfnamefont{C.}~\bibnamefont{Soul\'{e}}},\ }%
  \bibfield{title}{%
  \enquote{\bibinfo {title} {Analytic torsion and holomorphic determinant
  bundles i. bott-chern forms and analytic torsion},}\ }%
  \bibfield{journal}{%
  \bibinfo {journal} {Commun. Math. Phys.}\ }%
  \textbf{\bibinfo {volume} {115}},\ \bibinfo {pages} {49--78} (\bibinfo {year}
  {1988})\BibitemShut{NoStop}%
\bibitem{Bismut1988b}%
  \BibitemOpen
  \bibfield{author}{%
  \bibinfo {author} {\bibfnamefont{Jean-Michel}\ \bibnamefont{Bismut}},
  \bibinfo {author} {\bibfnamefont{Henri}\ \bibnamefont{Gillet}},\ and\
  \bibinfo {author} {\bibfnamefont{Christophe}\ \bibnamefont{Soul\'{e}}},\ }%
  \bibfield{title}{%
  \enquote{\bibinfo {title} {Analytic torsion and holomorphic determinant
  bundles},}\ }%
  \bibfield{journal}{%
  \bibinfo {journal} {Commun. Math. Phys.}\ }%
  \textbf{\bibinfo {volume} {115}},\ \bibinfo {pages} {79--126} (\bibinfo
  {year} {1988}),\
  \url{http://dx.doi.org/10.1007/BF01238854}\BibitemShut{NoStop}%
\bibitem{Bismut1988c}%
  \BibitemOpen
  \bibfield{author}{%
  \bibinfo {author} {\bibfnamefont{Jean-Michel}\ \bibnamefont{Bismut}},
  \bibinfo {author} {\bibfnamefont{Henri}\ \bibnamefont{Gillet}},\ and\
  \bibinfo {author} {\bibfnamefont{Christophe}\ \bibnamefont{Soul\'{e}}},\ }%
  \bibfield{title}{%
  \enquote{\bibinfo {title} {Analytic torsion and holomorphic determinant
  bundles},}\ }%
  \bibfield{journal}{%
  \bibinfo {journal} {Commun. Math. Phys.}\ }%
  \textbf{\bibinfo {volume} {115}},\ \bibinfo {pages} {301--351} (\bibinfo
  {year} {1988}),\
  \url{http://dx.doi.org/10.1007/BF01466774}\BibitemShut{NoStop}%
\bibitem{Henneaux1994}%
  \BibitemOpen
  \bibfield{author}{%
  \bibinfo {author} {\bibfnamefont{M.}~\bibnamefont{Henneaux}}\ and\ \bibinfo
  {author} {\bibfnamefont{C.}~\bibnamefont{Teitelboim}},\ }%
  \emph{\bibinfo {title} {Quantization of Gauge Systems}}\ (\bibinfo
  {publisher} {Princeton University Press},\ \bibinfo {year}
  {1994})\BibitemShut{NoStop}%
\bibitem{P.Griffiths1994}%
  \BibitemOpen
  \bibfield{author}{%
  \bibinfo {author} {\bibfnamefont{P.}~\bibnamefont{Griffiths}}\ and\ \bibinfo
  {author} {\bibfnamefont{J.}~\bibnamefont{Harris}},\ }%
  \emph{\bibinfo {title} {Principles of Algebraic Geometry}}\ (\bibinfo
  {publisher} {Wiley-Interscience},\ \bibinfo {year}
  {1994})\BibitemShut{NoStop}%
\bibitem{Blau1995}%
  \BibitemOpen
  \bibfield{author}{%
  \bibinfo {author} {\bibfnamefont{Matthias}\ \bibnamefont{Blau}}\ and\
  \bibinfo {author} {\bibfnamefont{George}\ \bibnamefont{Thompson}},\ }%
  \bibfield{title}{%
  \enquote{\bibinfo {title} {On diagonalization in map(m,g)},}\ }%
  \bibfield{journal}{%
  \Doi{10.1007/BF02104681}{\bibinfo {journal} {Commun. Math. Phys.}}\ }%
  \textbf{\bibinfo {volume} {171}},\ \bibinfo {pages} {639--660} (\bibinfo
  {year} {1995})\BibitemShut{NoStop}%
\bibitem{Hyun:1995is}%
  \BibitemOpen
  \bibfield{author}{%
  \bibinfo {author} {\bibfnamefont{Seungjoon}\ \bibnamefont{Hyun}}\ and\
  \bibinfo {author} {\bibfnamefont{Jae-Suk}\ \bibnamefont{Park}},\ }%
  \bibfield{title}{%
  \enquote{\bibinfo {title} {Holomorphic yang-mills theory and variation of the
  donaldson invariants},}\ }%
  \bibfield{journal}{%
  \bibinfo {journal} {hep-th/9503036}}%
   (\bibinfo {year} {1995})\BibitemShut{NoStop}%
\bibitem{Baulieu:1997nj}%
  \BibitemOpen
  \bibfield{author}{%
  \bibinfo {author} {\bibfnamefont{Laurent}\ \bibnamefont{Baulieu}}, \bibinfo
  {author} {\bibfnamefont{Andrei}\ \bibnamefont{Losev}},\ and\ \bibinfo
  {author} {\bibfnamefont{Nikita}\ \bibnamefont{Nekrasov}},\ }%
  \bibfield{title}{%
  \enquote{\bibinfo {title} {Chern-simons and twisted supersymmetry in various
  dimensions},}\ }%
  \bibfield{journal}{%
  \bibinfo {journal} {Nucl. Phys.}\ }%
  \textbf{\bibinfo {volume} {B522}},\ \bibinfo {pages} {82--104} (\bibinfo
  {year} {1998})\BibitemShut{NoStop}%
\bibitem{Witten1991a}%
  \BibitemOpen
  \bibfield{author}{%
  \bibinfo {author} {\bibfnamefont{Edward}\ \bibnamefont{Witten}},\ }%
  \bibfield{title}{%
  \enquote{\bibinfo {title} {On quantum gauge theories in two-dimensions},}\ }%
  \bibfield{journal}{%
  \Doi{10.1007/BF02100009}{\bibinfo {journal} {Commun. Math. Phys.}}\ }%
  \textbf{\bibinfo {volume} {141}},\ \bibinfo {pages} {153--209} (\bibinfo
  {year} {1991})\BibitemShut{NoStop}%
\bibitem{Moore2000}%
  \BibitemOpen
  \bibfield{author}{%
  \bibinfo {author} {\bibfnamefont{Gregory~W.}\ \bibnamefont{Moore}}, \bibinfo
  {author} {\bibfnamefont{Nikita}\ \bibnamefont{Nekrasov}},\ and\ \bibinfo
  {author} {\bibfnamefont{Samson}\ \bibnamefont{Shatashvili}},\ }%
  \bibfield{title}{%
  \enquote{\bibinfo {title} {Integrating over higgs branches},}\ }%
  \bibfield{journal}{%
  \Doi{10.1007/PL00005525}{\bibinfo {journal} {Commun. Math. Phys.}}\ }%
  \textbf{\bibinfo {volume} {209}},\ \bibinfo {pages} {97--121} (\bibinfo
  {year} {2000})\BibitemShut{NoStop}%
\bibitem{Witten1994}%
  \BibitemOpen
  \bibfield{author}{%
  \bibinfo {author} {\bibfnamefont{Edward}\ \bibnamefont{Witten}},\ }%
  \bibfield{title}{%
  \enquote{\bibinfo {title} {Supersymmetric yang-mills theory on a four
  manifold},}\ }%
  \bibfield{journal}{%
  \Doi{10.1063/1.530745}{\bibinfo {journal} {J. Math. Phys.}}\ }%
  \textbf{\bibinfo {volume} {35}},\ \bibinfo {pages} {5101--5135} (\bibinfo
  {year} {1994})\BibitemShut{NoStop}%
\bibitem{Blau2006}%
  \BibitemOpen
  \bibfield{author}{%
  \bibinfo {author} {\bibfnamefont{Matthias}\ \bibnamefont{Blau}}\ and\
  \bibinfo {author} {\bibfnamefont{George}\ \bibnamefont{Thompson}},\ }%
  \bibfield{title}{%
  \enquote{\bibinfo {title} {Chern-simons theory on s**1-bundles:
  Abelianisation and q- deformed yang-mills theory},}\ }%
  \bibfield{journal}{%
  \bibinfo {journal} {JHEP}\ }%
  \textbf{\bibinfo {volume} {05}},\ \bibinfo {pages} {003} (\bibinfo {year}
  {2006})\BibitemShut{NoStop}%
\end{thebibliography}
%

\end{document}